\documentclass[draftcls,12pt,onecolumn]{IEEEtran}

\usepackage{amsmath}
\usepackage{graphicx}
\usepackage[colorlinks,bookmarksopen,bookmarksnumbered,citecolor=red,urlcolor=red]{hyperref}

\newcommand\BibTeX{{\rmfamily B\kern-.05em \textsc{i\kern-.025em b}\kern-.08em
T\kern-.1667em\lower.7ex\hbox{E}\kern-.125emX}}

\begin{document}

\title{Power Allocation and Transmitter Switching for Broadcasting with Multiple Energy Harvesting Transmitters
\thanks{This research was supported by the National Natural Science Foundation of China (61162008), the Guangxi Natural Science Foundation (2013GXNSFGA019004), the Open Research Fund of Guangxi Key Lab of Wireless Wideband Communication \& Signal Processing (12103), the Director Fund of Key Laboratory of Cognitive Radio and Information Processing (Guilin University of Electronic Technology), Ministry of Education, China (2013ZR02), and the Innovation Project of Guangxi Graduate Education (YCSZ2012066).}}

\author{Fangfang Zhou, Hongbin Chen, Rong Yu, and Lisheng Fan
\thanks{F. Zhou and H. Chen are with the Key Laboratory of Cognitive Radio and Information Processing (Guilin University of Electronic Technology), Ministry of Education, Guilin 541004, China.}
\thanks{R. Yu is with the School of Automation, Guangdong University of Technology, Guangzhou, China.}
\thanks{L. Fan is with the Department of Electronic Engineering, Shantou University, Shantou, China.}}

\maketitle

\begin{abstract}
With the advancement of battery technology, energy harvesting communication systems attracted great research attention in recent years. However, energy harvesting communication systems with multiple transmitters and multiple receivers have not been considered yet. In this paper, the problem of broadcasting in a communication system with multiple energy harvesting transmitters and multiple receivers is studied. First, regarding the transmitters as a `whole transmitter' \cite{ck32}, the optimal total transmission power is obtained and the optimal power allocation policy in \cite{ck18} is extended to our system setup, with the aim of minimizing the transmission completion time. Then, a simpler power allocation policy is developed to allocate the optimal total transmission power to the data transmissions. As transmitter switching can provide flexibility and robustness to an energy harvesting communication system, especially when a transmitter is broken or the energy harvested by a transmitter is insufficient, a transmitter switching policy is further developed to choose a suitable transmitter to work whenever necessary. The results show that the proposed power allocation policy performs close to the optimal one and outperforms some heuristic ones in terms of transmission completion time. Besides, the proposed transmitter switching policy outperforms some heuristic ones in terms of number of switches.
\end{abstract}


\section{Introduction}
Recently, energy harvesting or rechargeable sensor networks emerge as a new paradigm of sensor networks, in which the nodes can harvest energy from nature \cite{ck32}--\cite{ck1}. Before this, sensor network nodes are powered by batteries with limited energy storage, which are hard to recharge or replace. Therefore, the key challenge is to save energy and prolong network lifetime while guaranteeing the application-specific performance. In contrast, the harvested energy relaxes the energy constraint thus extending network lifetime in energy harvesting sensor networks. However, the energy that can be harvested from the environment is unstable and varies over time. Hence, the harvested energy should be carefully utilized in order to maximize the utility of energy harvesting sensor networks.

A lot of excellent works on energy management in energy harvesting sensor networks have been done. For example, two-stage communication power management algorithms were proposed for maximizing the utility of energy harvesting sensors, considering the energy neutrality constraint, the fixed power loss effects of circuitry, and the battery inefficiency and its capacity \cite{ck7}. Energy allocation over source acquisition/compression and transmission for a single energy harvesting sensor was addressed that guarantee minimum average distortion while ensuring stability of the queue connecting source and channel encoders \cite{add1}. Discounted cost Markov decision process and reinforcement learning algorithms were applied to find optimal energy management policies to maximize the performance of a single energy harvesting sensor \cite{add4}. Through modeling the ambient energy supply by a two-state Markov chain and assuming a finite battery capacity, low-complexity transmission policies were proposed for a wireless sensor powered by an energy harvesting device \cite{msz}. Conditions for balancing a node's expected energy consumption with its expected energy harvesting capability in a uniformly-formed wireless sensor network were derived \cite{add2}. A stochastic Markov chain framework was proposed to characterize the interplay between the battery discharge policy and the irreversible degradation of the storage capacity \cite{add3}.

In addition, other energy harvesting communication systems were also investigated \cite{eh1}--\cite{eh8}. Specifically, many energy harvesting communication schemes have been designed toward the goal of minimizing the transmission completion time. For example, optimal packet scheduling in a point-to-point communication system was studied in \cite{ck16,ck17}. The goal is to adjust the transmission rate according to the data arrival and harvested energy, such that the time by which all packets are delivered is minimized. Transmission powers were optimized for a broadcasting communication system with an energy harvesting transmitter \cite{ck18,ck19,ck20}. The objective is to minimize the time by which all packets are sent to their destinations. In \cite{ck21}, this problem was further studied assuming a finite capacity battery. While \cite{ck18}, \cite{ck16}--\cite{ck21} studied packet scheduling over the additive white Gaussian noise (AWGN) channel, \cite{ck22} studied packet scheduling in a point-to-point communication system over fading channels. Except for the above representative works, the effects of multiple access channel, parallel and fading Gaussian broadcast channels, interference channel, time varying channels, wireless energy transfer, and packet arrivals during transmission were also taken into account \cite{ck23}--\cite{ck30}.

The earlier works \cite{ck16}-\cite{ck22} mainly considered energy harvesting communication systems with only one transmitter. However, nowadays many communication systems are equipped with more than one transmitter. Therefore, it is necessary to study energy harvesting communication systems with multiple transmitters. In \cite{ck23}, optimal packet scheduling in a multiple access communication system with two energy harvesting transmitters was investigated. In \cite{ck24}, a communication system with an energy harvesting transmitter over parallel and fading Gaussian broadcast channels was studied. In \cite{ck25}, an optimal power allocation policy for a communication system with two energy harvesting transmitters over an interference channel was proposed. These works shed light on energy harvesting communication systems with multiple transmitters, but did not consider transmitter switching. In our opinion, transmitter switching can provide flexibility and robustness to an energy harvesting communication system, especially when a transmitter is unable to send data or the energy harvested from the environment is insufficient for data transmission. If this happens, other neighboring transmitters can turn to work and help the transmitter to proceed data transmission. To make the transmitter switching effective and decrease the switching overhead, a well-designed policy is essential to choosing the suitable transmitter to work.

Motivated by the above fact and lying on the earlier works \cite{ck32,ck18,ck33}, power allocation and transmitter switching for broadcasting in a communication system with multiple energy harvesting transmitters and multiple receivers are studied in this paper. Our target is to minimize the transmission completion time and to reduce the number of switches under the energy causality constraint. The contributions of this paper are summarized as follows: 1) The optimal total transmission power and the optimal power allocation policy in \cite{ck18} are rebuilt in the communication system with multiple energy harvesting transmitters; 2) A new power allocation policy is proposed which performs close to the optimal one but is simpler; 3) A new transmitter switching policy is proposed for the communication system with multiple energy harvesting transmitters and multiple receivers, which is more complex than the communication systems we studied before \cite{ck32,ck33}.

The remainder of this paper is organized as follows. In Section \ref{sec2}, the energy harvesting communication system with multiple transmitters and multiple receivers is described. In Section \ref{sec3}, the power allocation and transmitter switching policies are elaborated. Simulation results are presented in Section \ref{sec4} and some concluding remarks are given in Section \ref{sec5}.

\section{Energy harvesting communication system model\label{sec2} }
\begin{figure}
\centering
\includegraphics[width=2.9in]{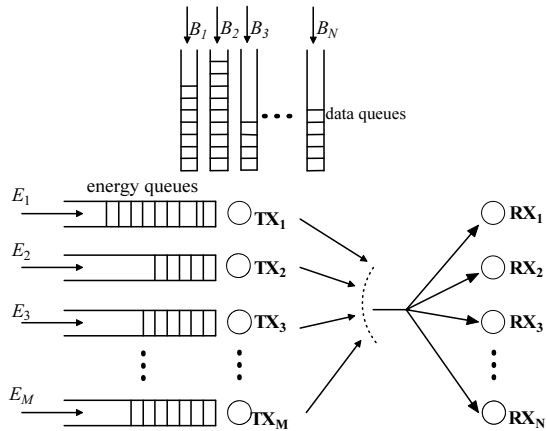}
\caption{Energy harvesting communication system with multiple transmitters and multiple receivers.}\label{tu1}
\end{figure}

We consider an energy harvesting communication system with multiple transmitters and multiple receivers, as shown in Figure~\ref{tu1}.
There are $M$ energy harvesting transmitters $\mbox{\rm TX}_1, \mbox{\rm TX}_2, \mbox{\rm TX}_3, \ldots, \mbox{\rm TX}_M$ and $N$ receivers $\mbox{\rm RX}_1, \mbox{\rm RX}_2, \mbox{\rm RX}_3, \ldots, \mbox{\rm RX}_N$. The energies arriving to the transmitters $E_1, E_2, E_3, \ldots, E_M$ are stochastic (both the arriving time and the amount are random) and independent of each other, while the data $B_1, B_2, B_3, \ldots, B_N$ are broadcasted by the transmitters in turn.  Here $B_n$ is the data to be sent to the receiver $\mbox{\rm RX}_n$ ($n=1, \ldots, N$). The energies arrive during the course of transmission while the data are given before transmission. For tractability, the arriving time and the amount of energies are assumed to be known at the beginning of transmission.  This system looks like a multi-input multi-output one. But we view the transmitters as a `whole transmitter' \cite{ck32} and focus on transmitter switching that can enhance flexibility and robustness of the system. The transmitters cooperate to send the data $B_1$, $B_2$, $B_3$, \ldots, $B_N$ to the corresponding receivers. Every time one of the transmitters $\mbox{\rm TX}_m$ will be active to broadcast data to the receivers. A transmitter switching policy will be designed to choose a suitable transmitter to work when the current working transmitter uses up its energy.

\begin{figure}
\centering
\includegraphics[width=3in]{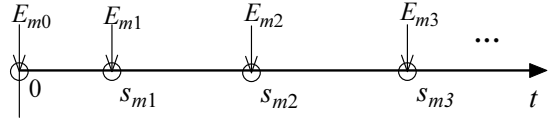}
\caption{Energy arriving process for the transmitter $\mbox{\rm TX}_m$.}\label{tu2}
\end{figure}

The energy arriving process for the transmitter $\mbox{\rm TX}_m$ is depicted in Figure~\ref{tu2}.
At the time $S_{mw}, w=1, 2, \ldots$, the amount of energy $E_{mw}$ arrives to $\mbox{\rm TX}_m$. $E_{m0}$ is the initial energy available in the battery of $\mbox{\rm TX}_m$ before the transmission starts. It is assumed that the batteries of the transmitters have infinite capacity and the harvested energy will not overflow.

Since there is only one transmitter sending data every time, the channel shown in Figure~\ref{tu1} is actually a broadcast channel. It is assumed that the chosen transmitter $\mbox{\rm TX}_m$ sends data to each receiver through an AWGN channel with different path losses. The signal received by the receiver $\mbox{\rm RX}_n$ can be represented by
 \begin{equation}\label{gso}
  y_{mn}=h_{mn}x+v_{mn},\quad m=1,\,\cdots,\,M;n=1,\,\cdots,\,N
\end{equation}
where $h_{mn}$ is the path loss between $\mbox{\rm TX}_m$ and $\mbox{\rm RX}_n$, $x$ is the transmitted signal, $v_{mn}$ is an AWGN with zero mean and variance $\sigma_{mn}^2$. Here $\sigma_{mn}^2=N_{mn}B_o$, where $N_{mn}$ is the noise power spectral density in the channel between $\mbox{\rm TX}_m$ and $\mbox{\rm RX}_n$, and $B_o$ is the bandwidth. It is assumed that all channels have the same bandwidth. Then, the capacity region for the broadcast channel is \cite{ck18}
\begin{align}\label{gs2}
&r_{mn}\leq B_o\log_2\bigg(1+\frac{P_n h_{mn}}{\sum_{j<n}P_j h_{mn}+N_{mn} B_o}\bigg),\nonumber\\
&\sum_{n=1}^{N}P_{n}=P_o.
\end{align}
where $r_{mn}$ is the transmission rate when $\mbox{\rm TX}_m$ sends data to $\mbox{\rm RX}_n$ with power $P_n$, $P_n$ is a portion of the total transmission power split to $\mbox{\rm RX}_n$, $P_o$ is the total transmission power. In the following, we analyze data transmission from the information-theoretic point of view. It is assumed that data transmission is always successful no matter which transmitter broadcasts data. Moreover, the transmitters will not send data that has been sent out. This can be coordinated by a central controller.

To minimize the transmission completion time (by which the given number of bits are delivered to their intended receivers), power allocation should be executed among the data transmissions under the energy causality constraint. The energy causality means that at any given time, the total amount of consumed energy must be no more than the total amount of harvested energy. Following our previous work in \cite{ck32}, we treat all the transmitters as a `whole transmitter' (\emph{we only care about the amount of bits sent to the receivers while the bits sent by which transmitter do not matter}) and find out the optimal total transmission power that achieves the maximum departure region \cite{ck18} for a given deadline (the dual problem of transmission completion time minimization). Then, we rebuild the optimal power allocation policy \cite{ck18} in our system setup. As the optimal power allocation policy needs the total transmission powers in all time slots to calculate the cut-off powers, we propose a simpler power allocation policy which only requires the total transmission power in the current time slot.  Since every time only one transmitter is active to send data, transmitter switching is unavoidable. However, more switching among transmitters will bring greater control overhead, even though the energy consumed for transmitter switching is relatively small. To reduce control overhead and to save energy, the number of switches should be as least as possible. Following our previous work in \cite{ck32,ck33}, we propose a transmitter switching policy to choose the suitable transmitter to send data with the principle of less number of switches. It should be emphasized that the turn of the working transmitters does not affect the transmission completion time. So we do power allocation first and then conduct transmitter switching. Note that with the optimal total transmission power at hand, transmitter switching can also be done before power allocation.

\section{Power allocation and transmitter switching policies\label{sec3}}
In this section, the power allocation policies and the transmitter switching policy will be presented.

\subsection{Optimal total transmission power and optimal power allocation policy\label{sec3.1}}
\begin{figure}
\centering
\includegraphics[width=3in]{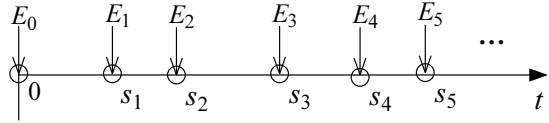}
\caption{Energy arriving process for the `whole transmitter'.}\label{tu3}
\end{figure}

With the aim of minimizing the transmission completion time, the optimal total transmission power was obtained in \cite{ck18}. Moreover, an optimal power allocation policy was derived for a broadcast communication system with an energy harvesting transmitter. Regarding the transmitters as a `whole transmitter', we record the energy arriving to the transmitters in chronological order, as shown in Figure~\ref{tu3}.
Here $E_0$ is the sum of the initial energy in the batteries. The `whole transmitter' harvests energy at the time instant $s_w$ with amount $E_w$. This energy can be harvested by an arbitrary transmitter that we do not need to know.

With the new energy arriving process, the optimal total transmission power will be calculated and some properties of the optimal power allocation policy will be referenced in the following:

First, from {\bf Lemma 1} we know that the total transmission power remains constant between two consecutive energy harvesting instants, that is, the total transmission power only changes at an energy harvesting time instant.

Second, from {\bf lemma 2} we get that the maximum departure region is a convex region. It means that there is one and only one optimal total transmission power.

Third, from {\bf Lemma 3} we derive the expression of the total transmission power as
\begin{align}\label{gss}
&i_l=\arg \min_{i_{l-1}<w<W} \{\frac{\sum_{j=i_{l-1}}^{w-1}E_j}{s_w-s_{i_{l-1}}}\},\nonumber\\
&P_{dl}=\frac{\sum_{w=i_{l-1}}^{i_l-1}E_w}{s_{i_l}-s_{i_{l-1}}}.
\end{align}
where $s_W$ is the transmission completion time and the energy arriving time before it is denoted as $s_{W-1}$, $P_{dl}$ is the optimal total transmission power for the `whole transmitter' $\mbox{\rm TX}_d$ over the interval $(s_{i_{l-1}},s_{i_l})$, $l=1,2,\cdots$.

After calculating the optimal total transmission power, we further split the power to the data transmissions. Without loss of generality, we rank all of the variances from $\sigma_{dn}^2$ as $\sigma_{1}^2\leq\sigma_{2}^2\leq \cdots \leq \sigma_{N}^2$ and denote the receiver corresponding to $\sigma_{n}^2$ as the $n$th receiver. Therefore, the first receiver is the strongest and the $N$th receiver is the weakest \cite{ck18}. From {\bf Lemma 4}, we know that there is a cut-off power for each of the strongest $N-1$ receivers, which are denoted as $P_{c1}, P_{c2}, \cdots, P_{c(N-1)}$. If the optimal total transmission power is below $P_{c1}$, all the power is allocated to the strongest receiver and the power allocated to the remaining $N-1$ receivers are zero. If the optimal total transmission power is higher than $P_{c1}$, the power allocated to the first receiver is $P_{c1}$. Then, we check whether the remaining power is below $P_{c2}$ or not. If the remaining power is higher than $P_{c2}$, the power $P_{c2}$ will be allocated to the second strongest receiver. Otherwise, all the remaining power will be allocated to the second strongest receiver and power will not be allocated to the remaining $N-2$ receivers. The rest can be done in the same manner.

From {\bf Corollary 1} of {\bf Lemma 4}, we know that the power for the data transmission to every receiver is either a non-negative constant sequence or an increasing non-negative sequence.

From {\bf Lemma 5} we know that with the optimal power allocation policy, all the data sent to the respective receivers must be finished at the same time.

\begin{figure}
\centering
\includegraphics[width=3in]{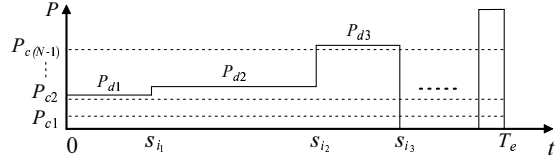}
\caption{Optimal total transmission power for the `whole transmitter'.}\label{tu4}
\end{figure}

With these properties and based on the results in \cite{ck18}, the optimal total transmission powers and the cut-off powers are obtained, which are plotted in Figure~\ref{tu4}.

\subsection{Proposed power allocation policy\label{sec3.2}}
\begin{figure}
\centering
\includegraphics[width=3in]{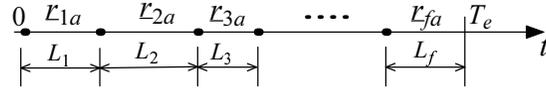}
\caption{Partitioning of time slots and corresponding transmission rates for the data $B_a$.}\label{tupr}
\end{figure}

Aiming at minimizing the transmission completion time, we propose an alternative power allocation policy that performs close to the optimal one but is simpler. The idea is that we heuristically make the transmission rates be proportional. This satisfies the properties mentioned in the previous subsection. To illustrate the proposed power allocation policy, we partition the total transmission time into several time slots according to Figure~\ref{tu4}. In every time slot, there is no transmitter switching and the transmission rate keeps constant. Next we derive the relationship between the amount of bits that to be transmitted and the transmission rates in all time slots. Take the data $B_a$ which corresponds to $\mbox{\rm RX}_a$ as an example. The partitioning of time slots and the corresponding transmission rates are shown in Figure~\ref{tupr}.
In the first time slot $L_1$, the transmission rate for $B_a$ is $\underline{r}_{1a}$. The rate during the next time slot is $\underline{r}_{2a}$, the transmission completion time is $T_e$, $f$ is the number of the time slots, and $f$ is equal to or greater than the number of switches (when the current working transmitter has energy left at the time the optimal total transmission power changes). The following equation can be easily obtained:
\begin{align}\label{gspr3}
B_a=\underline{r}_{1a}L_1+\underline{r}_{2a}L_2+\cdots+\underline{r}_{fa}L_f.
\end{align}
For notational simplicity, we continue the derivation at the system with three receivers, which are $\mbox{\rm RX}_1, \mbox{\rm RX}_2, \mbox{\rm RX}_3$. The derivation holds in the case of more receivers. The data to be delivered to the receivers are $B_1$, $B_2$, and $B_3$. The power allocated to $\mbox{\rm RX}_1, \mbox{\rm RX}_2$ and $\mbox{\rm RX}_3$ are $P_1$, $P_2$ and $P_3$, respectively. During every time slot, $P_1$, $P_2$ and $P_3$ are constant. The relationship between the total transmission power and $P_1$, $P_2$, $P_3$ is given by
\begin{align}\label{gspr4}
P_1+P_2+P_3=P_{dl}.
\end{align}
From (\ref{gspr3}), we get the following equations:
\begin{align}\label{gspr5}
&B_1=\underline{r}_{11}L_1+\underline{r}_{21}L_2+\cdots+\underline{r}_{f1}L_f,\nonumber\\
&B_2=\underline{r}_{12}L_1+\underline{r}_{22}L_2+\cdots+\underline{r}_{f2}L_f,\nonumber\\
&B_3=\underline{r}_{13}L_1+\underline{r}_{23}L_2+\cdots+\underline{r}_{f3}L_f.
\end{align}
We set $\frac{\underline{r}_{q1}}{\underline{r}_{q2}}=k_1$ and $\frac{\underline{r}_{q1}}{\underline{r}_{q3}}=k_2$ ($k_1$ and $k_2$ are constants). By substituting them in the first equation of (\ref{gspr5}), we can get that
\begin{align}\label{gspr6}
&B_1=k_1\underline{r}_{12}L_1+k_1\underline{r}_{22}L_2+\cdots+k_1\underline{r}_{f2}L_f=k_1B_2,\nonumber\\
&B_1=k_2\underline{r}_{13}L_1+k_2\underline{r}_{23}L_2+\cdots+k_2\underline{r}_{f3}L_f=k_2B_3.
\end{align}
Then, we obtain the relationship
\begin{align}\label{gspr7}
\frac{B_1}{\underline{r}_{q1}}=\frac{B_2}{\underline{r}_{q2}}=\frac{B_3}{\underline{r}_{q3}}.
\end{align}
Substituting (\ref{gs2}) into (\ref{gspr7}) and combining (\ref{gspr4}), the power allocation $P_1$, $P_2$, $P_3$ in every time slot can be obtained.

\subsection{Transmitter switching policy\label{sec3.3}}
In this subsection, a transmitter switching policy for choosing the suitable transmitter to work will be presented. With the optimal total transmission power and the allocated powers at hand, the transmission completion time can be determined. For a given transmission completion time $T_e$, the following propositions are introduced.

{\bf Proposition 1:} The transmitter which harvests energy at its last energy harvesting time before the transmission completion time will turn to work as long as the working transmitter uses up its energy.

\begin{figure}
\centering
\includegraphics[width=3in]{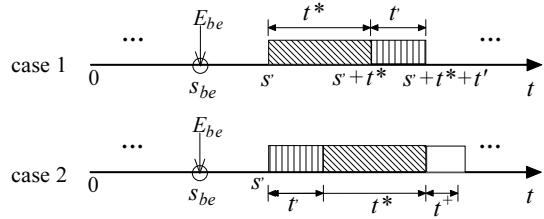}
\caption{Two cases of working order of the full transmitter and the partial transmitter.}\label{tu5}
\end{figure}

Proof: The transmitter which harvests energy at its last energy harvesting time before the transmission completion time is named as \emph{full transmitter} (which has finished energy harvesting) and the other transmitters are called \emph{partial transmitters}. We assume that the current working transmitter is $\mbox{\rm TX}_a$ and it sends data with power $P_{dl}$ (as transmitter switching is not affected by power allocation). Moreover, the optimal total transmission power keeps constant over several switches (if there is only one switch during the time the optimal total transmission power keeps constant, the following analysis still holds). The full transmitter $\mbox{\rm TX}_b$ harvests the last energy at the time instant $s_{be}$ with amount $E_{be}$, as shown in Figure~\ref{tu5}.
The total amount of energy available for $\mbox{\rm TX}_b$ is denoted by $E_{bo}$ (energies harvested by $\mbox{\rm TX}_b$ at earlier energy harvesting time may not be used up). The next switching instant is denoted by $s'$. For clarity, we still take the system with three transmitters as an example. The analysis can be easily extended to the system with more transmitters. The partial transmitter is denoted by $\mbox{\rm TX}_c$ with the amount of energy $E_{co}$ at the time instant $s'$. There are two possible cases which transmitter should work first between $\mbox{\rm TX}_b$ and $\mbox{\rm TX}_c$.

In Case 1, we assume that the partial transmitter $\mbox{\rm TX}_c$ works first. The amount of energy harvested by $\mbox{\rm TX}_c$ during the interval $(s', s'+t^\ast)$ is denoted by $E_{c1}$, and $t^\ast=\frac{E_{co}+E_{c1}}{P_{dl}}$. At the time instant $s'+t^\ast$, $\mbox{\rm TX}_b$ turns to work. The length of the working time slot for $\mbox{\rm TX}_b$ is $t'=\frac{E_{bo}}{P_{dl}}$. At the time instant $s'+t^\ast+t'$, another transmitter turns to work.

In Case 2, we assume that the full transmitter $\mbox{\rm TX}_b$ works first. At the time instant $s'+t'$, $\mbox{\rm TX}_c$ turns to work. During the interval $(s'+t^\ast, s'+t'+t^\ast+t^+)$, the amount of energy harvested by $\mbox{\rm TX}_c$ is $E_{c1'}$, and $t^+=\frac{E_{c1'}}{P_{dl}}$. It is easy to check that $t^+\geq0$. At the time instant $s'+t^\ast+t'+t^+$, another transmitter turns to work.

The length of the working time slots with twice switches in Case 2 must be longer than or equal to the one in Case 1. For a given transmission completion time $T_e$, the longer working time slot per switch will bring less number of switches. Hence the full transmitter should work first.

{\bf Proposition 2:} If there is more than one full transmitter, the working order of the full transmitters does not affect the number of switches.

Proof: When there are more than one full transmitter, we let all of them working earlier than the partial transmitters based on {\bf Proposition 1}. This prolongs the energy harvesting time for the partial transmitters before they use up their energies. As there is no energy arriving to the full transmitters, which full transmitter turns to work first has no influence on the working time slots for the remaining full transmitters. So the working order of the full transmitters has no effect on the partial transmitters, also does not affect the number of switches.

{\bf Proposition 3:} When no full transmitter exists in the system, the transmitter with the maximum amount of energy available should work first.

Proof: Greater amount of energy brings longer working time with the same transmission power. For a given transmission completion time $T_e$, longer working time leads to less number of switches.

\begin{figure}
\centering
\includegraphics[width=3in]{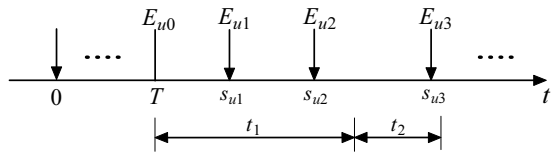}
\caption{The use of harvested energy for a partial transmitter $\mbox{\rm TX}_u$.}\label{tu6}
\end{figure}

With the above propositions, the suitable transmitter can be found. To help understanding the use of harvested energy, we take a partial transmitter $\mbox{\rm TX}_u$ as an example, as shown in Figure~\ref{tu6}.
The use of harvested energy for other transmitters are similar to $\mbox{\rm TX}_u$. In Figure~\ref{tu6}, $E_{u0}$ is the amount of energy available in the battery of $\mbox{\rm TX}_u$ at the present time $T$. At the time instant $s_{uw}$, $\mbox{\rm TX}_u$ harvests energy with amount $E_{uw}$. With the transmission power $P_{dl}$, the amount of energy $E_{u0}$ can make the transmitter work for a time slot $t_1$, where $t_1=E_{u0}/P_{dl}$. During this time slot, if there is new energy arriving, it will be harvested by $\mbox{\rm TX}_u$ and put into use before switching. For example, two energies $E_{u1}$ and $E_{u2}$ can be harvested before switching. Then, the new harvested energy $E_{u1}+ E_{u2}$ will be used to send data for a new time slot $t_2=\frac{E_{u1}+ E_{u2}}{P_{dl}}$. Until the time instant $T+ t_1+ t_2$, if there is new energy arriving, it will be harvested and used for keeping $\mbox{\rm TX}_u$ work; otherwise, if there is no new energy arriving, at the time instant $T+ t_1+ t_2$ another transmitter will turn to work.

\section{Simulation results\label{sec4}}
Numerical simulations are conducted to demonstrate the power allocation policies and the transmitter switching policy. First, the proposed power allocation policy is compared with the optimal power allocation policy. Then, the proposed power allocation policy is compared with some heuristic power allocation policies. Finally, the proposed transmitter switching policy is compared with some heuristic transmitter switching policies.

\subsection{Comparison with the optimal power allocation policy\label{secs1}}
We take the energy harvesting communication system with $M=3$ and $N=3$ as an example. The length between two consecutive energy arriving time for $\mbox{\rm TX}_1$, $\mbox{\rm TX}_2$ and $\mbox{\rm TX}_3$ obeys exponential distribution with parameters $\lambda_1=0.01$, $\lambda_2=0.1$ and $\lambda_3=1$, respectively. The amount of harvested energy $E_{mw}$ (mJ) obeys uniform distribution in the interval (0, 0.01), (0, 0.02) and (0, 0.03), respectively. Note that there is no actual model of the distributions of the stochastic energy arriving time and amount of arrived energy yet. We adopt these distributions just for exposition purpose. The analysis in the previous section does not depend on the distributions. The bits to be sent to $\mbox{\rm RX}_1$, $\mbox{\rm RX}_2$ and $\mbox{\rm RX}_3$ are $B_1$=70 bit, $B_2$=20 bit and $B_3$=10 bit, respectively. The three transmitters have the same channel parameters as follows: the bandwidth $Bo$=1 MHz; the path loss between $\mbox{\rm TX}_m$ and $\mbox{\rm RX}_1$, $\mbox{\rm RX}_2$, $\mbox{\rm RX}_3$ are $h_{m1}$=100 dB, $h_{m2}$=101 dB, $h_{m3}$=102 dB, respectively, $m=1, 2, 3$; the noise power spectral density is $N_{mn}=10^{-19}$ W/Hz, $m=1, 2, 3$, $n=1, 2, 3$. The transmission rates can be written as follows:
\begin{align}
&r_{m1}=\log_2\bigg(1+\frac{P_1}{10^{-3}}\bigg) \mbox{\rm Mbps},\nonumber\\
&r_{m2}=\log_2\bigg(1+\frac{P_2}{P_1+10^{-2.9}}\bigg) \mbox{\rm Mbps},\nonumber\\
&r_{m3}=\log_2\bigg(1+\frac{P_3}{P_1+P_2+10^{-2.8}}\bigg) \mbox{\rm Mbps}.
\end{align}

\begin{figure}
\centering
\includegraphics[width=3in]{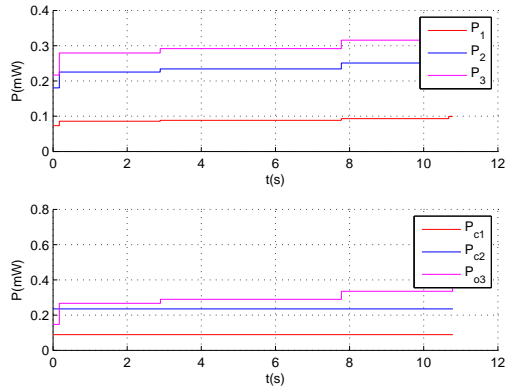}
\caption{Power allocation under the proposed policy (up) and the optimal policy (down).}\label{tudb}
\end{figure}

According to the above simulation parameters, we can get the optimal total transmission powers of the `whole transmitter' as $P_{d1}$=0.4712 mW, $P_{d2}$=0.5910 mW, $P_{d3}$=0.6139 mW, $P_{d4}$=0.6593 mW and $P_{d5}$=0.7263 mW. The corresponding time instants are $s_{i_1}$=0.1691 s, $s_{i_2}$=2.8973 s, $s_{i_3}$=7.7806 s, $s_{i_4}$=10.7788 s and $s_{i_5}$=10.7906 s. With the proposed power allocation policy, until the time instant 10.788761418 s, 1200 times of harvested energy is consumed by the system, the number of switches is 47, and all the bits are delivered to their intended receivers. We plot the allocated powers in the upper panel of Figure~\ref{tudb}. With the proposed power allocation policy, the power $P_1, P_2, P_3$ remain constant during a time slot and increase at the time instants $s_{il}$. We also plot the allocated powers under the optimal power allocation policy in the lower panel of Figure~\ref{tudb}. The transmission completion time of the optimal power allocation policy is 10.788513518 s. The power allocated to $\mbox{\rm RX}_1$ is a constant $P_{c1}$=0.0888 W, the power allocated to $\mbox{\rm RX}_2$ is also a constant $P_{c2}$=0.2354 W, and the remaining power $P_{o3}=P_{dl}-P_{c1}-P_{c2}$ is allocated to $\mbox{\rm RX}_3$. Because the optimal total transmission power is a constant or an increasing sequence, $P_{o3}$ changes simultaneously with $P_1$, $P_2$ and $P_3$. Even though the transmission completion time under the proposed power allocation policy is $2.4790\times10^{-4}$ s longer than the one under the optimal power allocation policy, the relative deviation is 0.04\%, which can be neglected.

\begin{figure}
\centering
\includegraphics[width=3in]{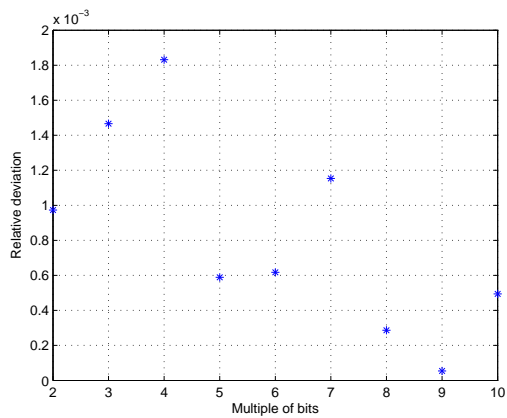}
\caption{Effect of multiple of bits on relative deviation.}\label{eab1}
\end{figure}

Moreover, we simulate the effect of multiple of bits on the relative deviation under the power allocation policies, as shown in Figure~\ref{eab1}.
The base of bits are $B_1$=7 bit, $B_2$=5 bit and $B_3$=2 bit. It is observed that the increase of number of bits may not enlarge the relative deviation, though it prolongs the transmission completion time. Moreover, the relative deviations are small for a moderate amount of bits.

\subsection{Comparison with heuristic power allocation policies\label{secs2}}
To show the advantage of the proposed power allocation policy, we compare it with some heuristic power allocation policies as follows.
\begin{enumerate}
\item {\bf Equal Power} (EP) policy:  in this policy, the optimal total transmission power is equally allocated to the three receivers. When the data transmission intended to a receiver is completed, that receiver is not involved in power allocation.
\item {\bf Data Ratio} (DR) policy:  the proposed power allocation policy allocates the optimal total transmission power according to the ratio of the amount of bits that to be transmitted and the transmission rate in every time slot. In this policy, the optimal total transmission power is allocated according to the ratio of the amount of bits that to be transmitted and the allocated powers in every time slot, which is
\begin{align}
\frac{B_1}{P_1}=\frac{B_2}{P_2}=\cdots=\frac{B_N}{P_N}.
\end{align}
 When the data transmission intended to a receiver is completed, that receiver is not involved in power allocation.
\item {\bf Remaining Data Ratio} (RDR) policy:  in this policy, the optimal total transmission power is allocated according to the ratio of the remaining bits and the allocated powers in every time slot, which is
\begin{align}
\frac{B_1-B_{o1}}{P_1}=\frac{B_2-B_{o2}}{P_2}=\cdots=\frac{B_N-B_{oN}}{P_N}.
\end{align}
where $B_{on}$ is the number of bits that has been sent to $\mbox{\rm RX}_n$ at the previous switching time instant. When the data transmission intended to a receiver is completed, that receiver is not involved in power allocation.
\end{enumerate}

In this subsection, we set $B_1$=15 bit, $B_2$=10 bit and $B_3$=7 bit. The other parameters are same as those in Subsection \ref{secs1}. Recall that the energy harvesting processes are stochastic. We take 1000 independent runs for the same setting and get the average transmission completion time, which are listed in Table \ref{tab1}.
\begin{table}
 \centering
\caption{\label{tab1}Average transmission completion time under the power allocation policies.}
 {\small\begin{tabular}{|p{1.4cm}|c|c|c|c|c|c}\hline
policy & average transmission completion time \\
\hline EP & 11.93 \\
\hline DR & 6.75 \\
\hline RDR & 6.20 \\
\hline Proposed & 3.46 \\\hline
\end{tabular}}
\end{table}
The proposed power allocation policy leads to the least average transmission completion time among the policies. The RDR policy allocates the power according to the remaining bits in time, which guarantees that all the data transmissions are completed nearly at the same time. Hence, the average transmission completion time under this policy is the second least. However, it is nearly double of the average transmission completion time under the proposed policy. The EP and the DR policies allocated the power in a fixed manner, which cannot guarantee the data transmissions are completed at the same time or nearly the same time. Thus, these policies lead to longer average transmission completion time.

\subsection{Comparison with heuristic transmitter switching policies\label{secs3}}
In this part, we compare the proposed transmitter switching policy with some heuristic ones under the proposed power allocation policy. The simulation parameters are the same as those in Subsection \ref{secs2}. The heuristic transmitter switching policies are given as follows.

\begin{enumerate}
\item {\bf Energy Minimum} (EM) policy:  in this policy, at every switching time instant, we choose the transmitter with the minimum energy to work.
\item {\bf Fixed Order 123} (FO123) policy:  in this policy, we let the order of switching be fixed: $\mbox{\rm TX}_1$ works first. When it uses up its energy, $\mbox{\rm TX}_2$ turns to work. $\mbox{\rm TX}_3$ works at last. When $\mbox{\rm TX}_3$ uses up its energy, a new turn starts again.
\item {\bf Fixed Order 132} (FO132) policy:  This policy is similar to the previous policy but the turn of switching changes, that is, $\mbox{\rm TX}_3$ works secondly and $\mbox{\rm TX}_2$ works at last.
\item {\bf Stochastic Switching} (SS) policy:  in this policy, when a transmitter uses up its energy, we choose another transmitter to work randomly.
\end{enumerate}
We take 10000 independent runs and get the average number of switches, which are listed in Table \ref{tab2}.
\begin{table}
 \centering
\caption{\label{tab2}Average number of switches under the transmitter switching policies.}
 {\small\begin{tabular}{|p{1.4cm}|c|c|c|c|c|c}\hline
policy & average number of switches \\
\hline Proposed & 18.43 \\
\hline EM & 20.33 \\
\hline FO123 & 26.08 \\
\hline FO132 & 26.09 \\
\hline SS & 46.39 \\\hline
\end{tabular}}
\end{table}
It is seen that the proposed policy leads to the least average number of switches among the policies. The EM policy chooses the transmitter with minimum energy, which means that each working time slot is short. Therefore, the number of switches under it must be greater than the one under the proposed policy. Both FO123 and FO132 policies have fixed switching order. Thus, they nearly attain the same average number of switches. The SS policy randomly chooses a transmitter to work, which brings the largest average number of switches. These three heuristic policies do not consider the amount of energy in the battery of transmitters. Their performances must be worse than that under the proposed policy.

\section{Conclusion\label{sec5}}
The problem of broadcasting in a communication system with multiple energy harvesting transmitters and multiple receivers has been discussed. To minimize the transmission completion time, we view the transmitters as a `whole transmitter', then calculate the optimal total transmission power and reiterate the optimal power allocation policy \cite{ck18} in our system setup. Moreover, to reduce the complexity of power allocation, a simpler power allocation policy is developed which nearly attains the same transmission completion time with the optimal one and leads to less transmission completion time than some heuristic ones. To enhance the flexibility and robustness of the system, a transmitter switching policy is further developed which leads to less number of switches than some heuristic ones.

\end{document}